\documentclass[12pt]{article}

\newcommand{\remove}[1]{}

\newcommand{\lae}{\stackrel{<}{\sim}}
\newcommand{\gae}{\stackrel{>}{\sim}}

\newcommand{\MOD}{{\cal{R}}}

\begin{document}

\begin{titlepage}

\title{Limitations of \\ B-meson mixing bounds on technicolor theories}

\author{Elizabeth H. Simmons\thanks{e-mail address: simmons@bu.edu} \\
Department of Physics, Boston University, \\ 590 Commonwealth Ave., 
Boston MA  02215 \\
Physics Department, Harvard University, Cambridge MA, 02138}
 
\date{Nov. 2, 2001}

\maketitle
\thispagestyle{empty}

  \begin{picture}(0,0)(0,0)
    \put(400,300){BUHEP-01-28}
    \put(400,280){HUPT-01/A054}
  \end{picture}
  \vspace{24pt}


\begin{abstract}
  Recent work by Burdman, Lane, and Rador has shown that $B\bar{B}$
  mixing places stringent lower bounds on the masses of topgluons and
  $Z^\prime$ bosons in classic topcolor-assisted technicolor (TC2)
  models.  This paper finds analogous limits on the $Z^\prime$ bosons
  of flavor-universal TC2 and non-commuting extended technicolor
  models, and compares the limits with those from precision
  electroweak measurements.  A discussion of the flavor structure of
  these models (contrasted with that of classic TC2) shows that
  B-meson mixing is a less reliable probe of these models than of
  classic TC2.

\end{abstract}

\end{titlepage}

\newpage
\renewcommand{\thepage}{\arabic{page}}
\setcounter{page}{1}

The flavor problem remains a challenge for dynamical models of mass
generation.  While technicolor \cite{technicolor} can provide appropriate
masses for the electroweak gauge bosons, explaining the masses and mixings of
the quarks and leptons is more difficult.  Extended technicolor models
\cite{extendedtc} postulate an enlarged gauge group coupling the quarks and
leptons to the technifermion condensate, enabling them to acquire mass.
However, the simplest models of this type tend to produce large
flavor-changing neutral currents.  Given the large value of the top quark's
mass and the sizable splitting between the masses of the top and bottom
quarks, it is natural to wonder whether $m_t$ has a different origin than the
masses of the other quarks and leptons.  A variety of dynamical models that
exploit this idea have been proposed, including topcolor-assisted technicolor
(TC2)\cite{topassist}, flavor-universal TC2 \cite{Popovic:1998vb}, and
non-commuting extended technicolor (NCETC) \cite{Chivukula:1994mn}.

A common feature of these models is that they extend one or more of the
standard model $SU(N)$ gauge groups to an $SU(N) \times SU(N)$ structure at
energies well above the weak scale.  The TC2 models have extended color and
hypercharge groups; NCETC has an extended weak gauge group. Spontaneous
breaking of the extended groups to their diagonal (standard model) subgroups
produces extra massive gauge bosons: a color-octet of topgluons plus a
$Z^\prime$ boson in TC2, a color-octet of colorons and a $Z^\prime$ boson in
flavor-universal TC2, and a trio of $W^\prime$ and $Z^\prime$ bosons in
NCETC.  In order to use their enlarged gauge groups to help explain the heavy
top quark mass, all of these models propose that some of the gauge groups
should be flavor non-universal, treating the third generation fermions
differently than those in the first and second generations.  Accordingly, the
topgluons and $Z^\prime$ of classic TC2, the $Z^\prime$ of flavor-universal
TC2, and the extra weak bosons of NCETC all have flavor non-universal
couplings -- meaning that these gauge bosons can cause tree-level
flavor-changing neutral currents.

Recently, Burdman et al. \cite{Burdman:2001in} pointed out that $B_d
\bar{B}_d$ mixing provides strong lower bounds on the masses of the topgluon
and $Z^\prime$ bosons of classic TC2 models.  They briefly mentioned that
their work should constrain flavor-universal TC2 models, but did not present
specific results for those models.  This paper explicitly extends the
analysis to study both flavor-universal TC2 and non-commuting ETC.  We find
that the limits on the gauge bosons of these models are numerically much
weaker than those for classic TC2 models and are comparable to limits from
precision electroweak observables.  In addition, we show that the flavor
structure of these models differs from that of classic TC2 in ways that
render limits from B-meson mixing less generally applicable.

Exchange of a heavy gauge boson $G$ with non-universal
couplings of the type discussed above gives rise to the
following $\vert \Delta B\vert = 2$ interaction at low energies:
\pagebreak
\begin{eqnarray}
{\cal H}_G = \frac{2\pi\alpha_G}{M^2_G} \sum_{\lambda_1,\lambda_2 = L,R} 
\left(D^*_{\lambda_1 bb} D_{\lambda_1 bd_i} 
  [A^b_{G\lambda_1} - A^d_{G\lambda_1}]
  (\bar{b}_{\lambda_1}\gamma^\mu \Gamma^G d_{\lambda_1})\right) \times \nonumber \\
\left(D^*_{\lambda_2 bb} D_{\lambda_2 bd_i} 
  [A^b_{G\lambda_2} - A^d_{G\lambda_2}]]
  (\bar{b}_{\lambda_2}\gamma_\mu \Gamma^G d_{\lambda_2})\right) 
+ h.c.
\label{extensio}
\end{eqnarray}
The mass of the gauge boson is $M_G$ and the fine structure constant
of the relevant Standard Model gauge group is $\alpha_G$.  The matrix
$\Gamma^G$ is $\lambda^a/2$ [$T^a$,1] if $G$ comes from an extended
SU(3) [SU(2), U(1)] gauge group.  The factor $A^b{G\lambda_i}$
($A^d{G\lambda_i}$) is the full strength (modulo the gauge coupling
$\sqrt{4\pi\alpha_G}$) with which boson $G$ couples to a
$\lambda_i$-handed $b$ ($d$ or $s$) quark through
the $SU(N)_1$ and $SU(N)_2$ bosons and any bosons that mix with them.  The
matrices $D_{\lambda_i}$ represent the mixing between
$\lambda_i$-handed down-type quarks. Equation (\ref{extensio}) is a
slight generalization of equations (5) and (6) of
ref. \cite{Burdman:2001in}; the additional terms included here are
numerically irrelevant for the results of Burdman et al., but will be
needed in our analysis.

As discussed in ref. \cite{Burdman:2001in}, the primary effect of the
topgluon from classic TC2 on B-meson mixing comes from its exchange
among left-handed gauge-eigenstate $b$-quarks.  Since $b$ is an
$SU(3)_1$ triplet and an $SU(3)_2$ singlet, while $d$ (and $s$) is
just the opposite, one finds $A^b_{CL} = \cot\theta_C$ and $A^d_{CL} =
-\tan\theta_C$ in eq. (\ref{extensio}).  The main contribution
from the Z' comes from the $U(1)_{1,2}$ charges ($Y_{1,2}$) of the left-handed
down-type quarks;  e.g., $A^b_{YL} = Y^b_{1L}
\cot\theta_Y - Y^b_{2L}\tan\theta_Y$. The combined contribution to the
$B^0_L B^0_S$ mass difference is given \cite{Buras:1998fb} by $\Delta
M_{B_d} = 2 \vert M_{12} \vert_{TC2}$
\begin{eqnarray}
2 (M_{12})_{TC2} = \frac{4\pi}{3}  \eta_B
      M_{B_d} f^2_{B_d} B_{B_d} (D^*_{Lbb} D_{Lbd})^2 \MOD_{TC2} 
\label{onearr}\\
\MOD_{TC2} =  \left[ \frac{\alpha_C \cot^2\theta_C (1 + \tan^2\theta_C)^2}{3
      M_C^2} + \frac{\alpha_Y \cot^2\theta_Y 
     ((Y^b_{1L} - Y^d_{1L}) - (Y^b_{2L} - Y^d_{2L})\tan^2\theta_Y)^2}{M_{Z'}^2}
      \right]
\label{twoarr}
\end{eqnarray}
where the dependence on gauge boson couplings and masses has been
collected in the expression $\cal{R}$ for later convenience.  In the
above equations, $\alpha_C$ ($\alpha_Y$) is the Standard Model fine
structure constant for color (hypercharge), $M_C$ ($M_{Z'}$) is the
topgluon ($Z^\prime$) mass, and $\theta_C$ ($\theta_Y$) is the mixing
angle between the two original color (hypercharge) groups.  Following
ref. \cite{Burdman:2001in}, we set the QCD radiative correction factor
for the LL product of color-singlet currents to $\eta_B = 0.55 \pm
0.01$ and use $f_{B_d} \sqrt{B_{B_d}} = (200 \pm 40)$ MeV
\cite{Buras:1998fb} to incorporate the $B_d$ meson decay constant and
bag parameter.

Ref. \cite{Burdman:2001in} argues that, in classic TC2 models, quark mixing
occurs almost exclusively in the left-handed down-quark sector of those
models and, in consequence, the factor $D^*_{Lbb} D_{Lbd}$ appearing in the
expression for $(M_{12})_{TC2}$ is approximately equal to $V^*_{tb} V_{td}$.
This means that the quark mixing factor appearing in $(M_{12})_{TC2}$ is
identical to that in the standard one-loop contribution from W exchange
(which is also present in ETC and TC2 models).  The two contributions to
neutral B-meson mixing therefore add constructively.  We will refer to this
situation as the ``constructive'' scenario.  To facilitate a comparison of
the limits obtained for different models, we will start by assuming that the
constructive scenario also applies in the models we are considering.  Then,
we will examine quark mixing in the flavor-universal TC2 and NCETC models in
more generality and discuss the implications.

We can find limits on the $Z^\prime$ bosons of the flavor-universal
TC2 and NCETC models by scaling from the results of
ref. \cite{Burdman:2001in} based on the $\MOD$ factors for the
different models.  To begin, we reproduce the value of $\MOD_{TC2}$
obtained in ref. \cite{Burdman:2001in}.  On the RHS of eq.
(\ref{twoarr}), the first term dominates, due to the large topgluon
gauge coupling and the anticipated similar sizes of the topgluon and
$Z^\prime$ masses.  Following ref.  \cite{Burdman:2001in}, we
approximate $\MOD_{TC2}$ by its first term, setting $\cot^2\theta_C
\approx 25$ and $\alpha_C$(1 TeV) = 0.093, and obtain
\begin{equation} 
\MOD_{TC2} \approx \frac{\alpha_C \cot^2\theta_C}{3 M_C^2} \approx
\frac{0.8}{M_C^2} .
\end{equation}
Ref. \cite{Burdman:2001in} showed that applying the experimental limit
$\Delta M_{B_d} = (3.11 \pm 0.11) \times 10^{-13}$ GeV \cite{PDBook}
to eq. (\ref{onearr}) yields a lower bound on the topgluon mass.
The value of the bound is $M_C \gae 3.1$ TeV if ETC contributes
significantly to the Kaon CP-violation parameter $\epsilon$ and $M_C
\gae 4.8$ TeV if it does not.

In the flavor-universal TC2 model \cite{Popovic:1998vb}, the gauge
group is the same as in classic TC2 \cite{topassist}: $G_{ETC} \times
SU(3)_1 \times SU(3)_2 \times SU(2) \times U(1)_1 \times U(1)_2$.  The
fermion charge assignments are different than in classic TC2: here,
all quarks are $SU(3)_1$ triplets and $SU(3)_2$ singlets.  The
color-octet of coloron bosons in the low-energy spectrum therefore
couples with equal strength ($A^b_C = A^d_C = \cot\theta_C$) to all
quarks and does not generate flavor-changing neutral currents (unlike
the topgluons of classic TC2).  The factor $\MOD$ does not have a
coloron contribution, only a contribution from the non-universal
$Z^\prime$ bosons (third-generation fermions feel $U(1)_1$ and the
others feel $U(1)_2$):
\begin{equation}
\MOD_{univ-TC2} =  \left [\frac{\alpha_Y \cot^2\theta_Y 
(Y^b_{1L} + Y^d_{2L}\tan^2\theta_Y)^2}{M_{Z'}^2}
\right] 
\label{utceq}
\end{equation}
As discussed in ref. \cite{Popovic:1998vb}, self-consistency in the
treatment of the more strongly-coupled $U(1)_1$ group (i.e., avoiding
the Landau pole) requires $\alpha_Y \cot^2\theta_Y \equiv \kappa_1
\lae 1$.  The benchmark flavor-universal model has standard-model
values for the fermion hypercharges, so that $Y^b_{1L} +
Y^d_{2L}\tan^2\theta_Y)^2 \approx 1/36$ (for small $\tan^2\theta_Y$).  Then
\begin{equation} 
\MOD_{univ-TC2} \approx \frac{0.028}{M_{Z'}^2} .
\end{equation}
The lower limit on $M_{Z'}$ of flavor-universal TC2 is roughly a
factor of 5 weaker than that on the topgluons of classic TC2. Scaling from
the results of ref. \cite{Burdman:2001in}, we find $M_Z' \gae 590$ GeV if ETC
does contribute to $\epsilon$ and $M_Z' \gae 910$ if it does not.  This is
comparable to the previous lower bounds on the $Z^\prime$ mass from precision
electroweak fits \cite{Chivukula:1996cc}, which range from 500 GeV to over 2
TeV depending on the value of $\kappa_1$.  It is stronger than those from LEP
II searches for anomalous four-fermion couplings \cite{Lynch:2001md}, which
are of order 400 GeV. 

The NCETC models \cite{Chivukula:1994mn} do not include non-standard
colored gauge bosons at all. In these models, it is the weak gauge
sector which is extended to $SU(2)_h \times SU(2)_l$.  Under the weak
sector, third-generation fermions transform as (2,1) and light
fermions as (1,2).  Because the $Z^\prime$ boson arises from the
mixing of two $SU(2)$ groups (with gauge coupling ratio $g_h / g_l
\equiv \tan\phi$), the values of the coupling factors in
eq. (\ref{extensio}) are $A^b_{WL} = -\frac12 {\tan\phi} + {\cal A}$
and $A^d_{WL} = \frac12 \cot\phi + {\cal A}$ where {\cal A} is a
flavor-universal term\footnote{Note that ${\cal A} = \sec^2\theta_W (-\frac12 +
\frac13\sin^2\theta_W)\cot\phi (M_W / M_{Z'})^2$ is rendered
negligible by the experimental limit $M_{Z'} \gae 400$ GeV and the
requirement $\sin^2\phi \leq 0.97$ \cite{Chivukula:1994mn} in order
that quarks not condense.} that cancels when the difference is taken.
Thus, $\MOD$ is
\begin{equation}
\MOD_{NCETC} =  \left [\frac{\alpha_W (\tan\phi + \cot\phi)^2}{4 M_{Z'}^2}
\right] \approx \frac{0.008}{\sin^2\phi \cos^2\phi M_{Z'}^2}\ .
\label{nceeq}
\end{equation}
When $\tan^2\phi \sim 1$, the lightest allowed mass for this
$Z^\prime$ boson will be about one-fifth that for for the topgluon of
classic TC2: 620 (960) GeV if ETC does (not) contribute to $\epsilon$.
For ``heavy case'' NCETC, where the same condensate supplies the
masses of the top quark and the electroweak bosons, this bound is far
weaker than existing limit $M_{Z'} \gae 1.5$ TeV from precision
electroweak data \cite{Chivukula:1996gu}.  For the opposite, ``light
case'' of NCETC, the precision electroweak data
\cite{Chivukula:1996gu} and LEP II compositeness searches
\cite{Lynch:2001md} allow the $Z^\prime$ bosons to be as light as
400-600 GeV when $\tan^2\phi \gae 1.85$ [i.e., $\sin^2\phi \gae
0.65$], but restrict the $Z^\prime$ to heavier masses for smaller
$\tan\phi$.  Hence the limit from B-meson mixing is weaker than that
from other sources for $\tan^2\phi \lae 1.85$, but could be comparable
or stronger for larger $\tan^2\phi$ (especially if ETC does not
contribute to $\epsilon$).  For example, if we set $\tan^2\phi =3$
[i.e., $\sin^2\phi = 0.75$] in eq. (\ref{nceeq}) and scale from the
results of ref.  \cite{Burdman:2001in}, the bound from FCNC is $M_{Z'}
\gae 650$ GeV if ETC does contribute to $\epsilon$ ($M_{Z'} \gae$ 1
TeV if it does not).

Having found the limits on the $Z^\prime$ masses in the constructive
scenario, we need to discuss the extent to which that scenario applies
to the flavor-universal TC2 and NCETC models.  By reviewing the
arguments and constraints which push flavor mixing into the $D_L$
sector in classic TC2, we can evaluate the extent to which this is
true of the other models.

In classic TC2 models, it is argued \cite{Burdman:2001in, Lane:1995gw}
that both the $U$ and the $D_R$ matrices are nearly
block-diagonal\footnote{Ref. \cite{Buchalla:1995dp} discusses the
alternative possibility that the $U_L$ and $D_L$ matrices have
triangular textures, leaving the $U_R$ and $D_R$ matrices essentially
unconstrained.  The triangular textures tend to weaken or eliminate
the B-meson mixing bound.}  (a 2x2 block for the two light generations
and a 1x1 block for the third generation), forcing essentially all the
quark mixing of the third generation with the other two into $D_L$.
This texture for the $U$ matrices essentially corresponds to having
$m_t$ arise from different physics than $m_u$ and $m_c$, a condition
that does obtain in both flavor-universal TC2 and NCETC models.
Separately, $D_R$ is constrained to be nearly block-diagonal in TC2 by
the requirement that the b-pions formed by the strong topcolor
dynamics not make unduly large contributions to $B\bar{B}$ mixing.
Kominis \cite{Kominis:1995fj} observed that agreement of the mixing
rate with experiment required
\begin{equation}
\frac{\vert D^*_{Lbd} D_{Rbb} D^*_{Rbd} D_{Lbb} \vert}{M_\pi^2} < 10^{-12}\ 
{\rm GeV}^{-2}\ .
\label{komeq}
\end{equation}
He pointed out that this inequality would be violated by approximately two
orders of magnitude if the elements of $D_L$ and $D_R$ were of order the
square root of CKM elements (for b-pion masses of order several hundred GeV).

In the case of NCETC models, there are no topcolor dynamics and no
b-pions to affect $B\bar{B}$ mixing.  In flavor-universal TC2 models,
the topcolor dynamics form a full set of q-pions, and a GIM mechanism
prevents tree-level FCNC \cite{Popovic:1998vb}.  For contributions at
one loop in flavor-universal TC2, the additional phase space
suppression factor of $1/16\pi^2$ alone would suffice to satisfy
eq. (\ref{komeq}) even for $D_L \sim D_R \sim \sqrt{CKM}$.  We
conclude that significant inter-generational mixing in $D_R$ is not
forbidden in either flavor-universal TC2 or NCETC models.

Once mixing is allowed in the right-handed down-quark sector, the
strict constraints of the constructive scenario evaporate.  In
eq. (\ref{onearr}), the factor $(D^*_{Lbb} D_{Lbd})^2$ entering the
contributions to $\Delta M_{B_d}$ from products of left-handed
currents can take on values much larger or much smaller than
$V_{tb}^*V_{td}$.  Moreover, right-handed fermion currents may now
also contribute to $B\bar{B}$ mixing, with a relative sign that is not
universally fixed.  One can see from eq. (\ref{extensio}) that the
contributions of a product of right-handed currents to $2\vert
M_{12}\vert$ would replace the factor $(D^*_{Lbb} D_{Lbd})^2$ in
eq. (\ref{onearr}) by $(D^*_{Rbb} D_{Rbd})^2$; a product of a left-
and a right-handed current would replace it by $(2 D^*_{Lbb} D_{Lbd}
D^*_{Rbb} D_{Rbd})$.  At the same time, in the context of
flavor-universal TC2, the hypercharge factors in $\MOD_{univ-TC2}$
(eq. (\ref{utceq})) would become $(Y^b_{1R} +
Y^d_{2R}\tan^2\theta_Y)^2$ for a product of right-handed currents or
$(Y^b_{1L} + Y^d_{2L}\tan^2\theta_Y)(Y^b_{1R} +
Y^d_{2R}\tan^2\theta_Y)$ for a mixed product.  In NCETC models,
because both $SU(2)$ groups are left-handed, the leading couplings of
right-handed quarks to the $Z^\prime$ are flavor-universal: $A^q_{WR}
= -Q_q \tan^2\theta_W \cot\phi\, (M_W/M_{Z'})^2$ where $Q$ is electric
charge.  Hence, if mixing in the right-handed down sector reduced the
size of $(D^*_{Lbb} D_{Lbd})^2$, right-handed currents could not
``pick up the slack'' and produce a net NCETC contribution to B-meson
mixing of the size predicted by the constructive scenario.  Due to the
uncertainties introduced when $D_R$ is no longer block-diagonal, it
becomes impossible to give a simple, universal size of the
contribution to $\Delta M_{B_d}$ in the whole class of NCETC or
flavor-universal TC2 models.

Finally, we note that the B-meson mixing limits discussed
here for NCETC models also generally apply to topflavor models
\cite{Li:nk,Muller:1996dj,Malkawi:1996fs} which share the same
extended electroweak gauge structure, but use Higgs bosons instead
of technicolor to break the electroweak symmetry.  Previous studies of
these models have noted that large contributions to B-meson mixing
are possible \cite{Lee:1998qq, Malkawi:1999sa} -- even though
electroweak data sets a lower bound of 1.7 TeV on the $Z^\prime$ boson
\cite{Malkawi:1999sa} -- but have not provided specific limits on
model parameters.

To summarize: In the flavor-universal topcolor-assisted technicolor and
non-commuting extended technicolor models, the constructive scenario of quark
mixing, in which only left-handed quark currents contribute and they do so
proportional to $(V_{tb}^* V_{td})^2$, is merely a useful benchmark, not a
requirement as in classic TC2 models.  Even if one assumes the constructive
scenario to apply, the resulting limits on gauge boson masses from $B\bar{B}$
mixing are similar in strength to more generally applicable limits based on
electroweak data.  We therefore conclude that $B\bar{B}$ mixing represents a
far less significant constraint on these models than on classic
topcolor-assisted technicolor.

\centerline{\bf Acknowledgments}

Thanks are due to R.S. Chivukula, K. Lane, G. Burdman, and C. Hill for
useful conversations and comments on the manuscript.  The hospitality
of the Aspen Center for Physics during the completion of this work is
gratefully acknowledged.  {\em This work was supported in part by the
Department of Energy under grant DE-FG02-91ER40676 and by the National
Science Foundation under grant PHY-0074274.}


\end{document}